\documentclass[%
 reprint,
 amsmath,amssymb,
 aps,
]{revtex4-2}

\usepackage{graphicx}
\usepackage{dcolumn}
\usepackage{bm}

\usepackage[usenames]{color}
\begin{document}

\preprint{APS/123-QED}

\title{Effective interfacial Dzyaloshinskii-Moriya interaction and skyrmion stabilization in ferromagnet/paramagnet and ferromagnet/superconductor hybrid systems}

\author{M.A. Kuznetsov}
 \email{kuznetsovm@ipmras.ru}

 \author{A.A. Fraerman}%
 \email{andr@ipmras.ru}
\affiliation{%
 Institute for Physics of Microstructures, Russian Academy of Sciences, Akademicheskaya St. 7,
Nizhny Novgorod 607680, Russian Federation
}%


\author{K.R. Mukhamatchin}%
 \email{mykamil@yandex.ru}
\affiliation{
 Lobachevsky State University of Nizhny Novgorod, Faculty of Physics, 23 Gagarin Av., Nizhny Novgorod 603950, Russian Federation
}%



\begin{abstract}
It is shown that a term in the form of Dzyaloshinskii-Moriya interaction (DMI) contributes to the free energy of a ferromagnetic (FM) film on a paramagnetic (PM) (an FM above the critical temperature, $T_c$) or superconducting (SC) substrate occurring in the London limit. This contribution results from magnetostatic interaction between the film and substrate under which the substrate affects FM magnetization back via its magnetic field produced by magnetization inhomogeneity in the film. Strikingly, in the FM/PM system this effective DMI stabilizes chiral magnetic textures, e.g., magnetic skyrmions (MSk´s) of the Néel-type, which is in contrast to that in the FM/SC one. A strong temperature sensitivity of the effective DMI allows for tuning the coupling between the FM film and PM or SC substrate and thus controlling the MSk radius in FM/PM. 
\end{abstract}

\maketitle


\section{\label{I}INTRODUCTION}

Magnetic skyrmions (MSk´s) are chiral spin textures with vortex configuration, the existence of which was predicted earlier \cite{Bog}, are of potential interest in the field of information storage, processing devices, and neuromorphic computing \cite{rev1,rev2,rev3,rev4}. As well known, such topologically protected structures result from Dzyaloshinskii–Moriya interaction (DMI) which causes noncollinear ordering of magnetic moments in media without an inversion center \cite{Dz,Mor}. An example of a medium with stable MSk´s, in which the inversion symmetry is broken by the interface, is a ferromagnet (FM)/heavy metal (HM) bilayer, where the HM provides the DMI in the system of FM magnetic moments \cite{DMI1,DMI2,DMI3,DMI4,DMI5}. This DMI removes the chiral degeneracy, thus mediating nonreciprocal spin-wave propagation \cite{DMISW1,DMISW2} and formation of chiral domain walls \cite{Bog,DMI2,DW}. It is also of interest that the chiral symmetry can be broken by magnetostatic interaction between the FM film and polarizable substrate (paramagnet (PM) \cite{MS1,MS2,MS3} or a superconductor (SC) \cite{MS2,MS3,MS4,MS5}). One consequence of this coupling is spin-wave nonreciprocity \cite{MS3,MSSW1,MSSW2,MSSW3,MSSW4,MSSW5,MSSW6,MSSW7,MSSW8} in FM/PM \cite{MS3,MSSW3,MSSW4,MSSW5,MSSW6,MSSW7,MSSW8} and FM/SC \cite{MS3,MSSW1,MSSW2} systems. Apart of these studies it was found that the magnetostatic interaction in similar systems contributes to the stabilization of chiral magnetic textures \cite{stab1,stab2,stab3,stab4}. It has also been shown that the SC has a significant effect on the MSk \cite{SC1,SC2,SC3,SC4,SC5,SC6,SC7}. The latter studies point out that MSk´s can be stabilized in an FM film on a polarized substrate due to magnetostatic interaction.

\begin{figure}[h!!]
\includegraphics{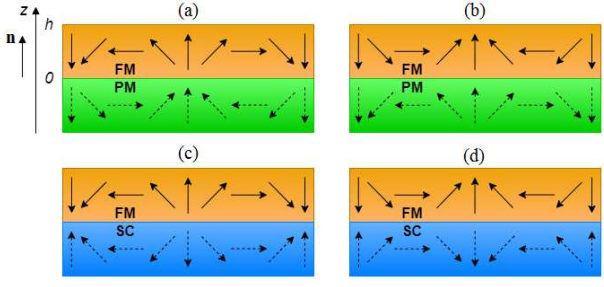}
\caption{\label{fig:1} Schematic for a magnetic spiral (skyrmion) of the Néel-type (solid arrows) and its images (dashed arrows) in FM/PM (a, b) and FM/SC (c, d). The right ($\psi=0$, $\sigma=1$) (a, c) and left ($\psi=\pi$, $\sigma=-1$) (b, d) rotations of the solid arrows lead to the left and right rotations of the dashed ones, respectively.}
\end{figure}

In this paper, we theoretically study FM/PM and FM/SC hybrid systems, where the PM is an FM above $T_c$ and SC in FM/SC is considered in the London limit below $T_c$, where $T_c$ is the Curie temperature in PM or the critical temperature for the normal metal-SC phase transition. We show that the effective DMI in the systems under study arises as an additional term in the energy of the FM film,
\begin{equation}
F_\text{DMI}^\text{eff}\propto\int_{-\infty}^{+\infty} D_\text{eff}(q)(\mathbf{n}\times\mathbf{q})\cdot[\mathbf{M(-q)}\times\mathbf{M(q)}]\,d\mathbf{q}
\label{eq:1},
\end{equation}
where $D_\text{eff}$ is the effective DMI constant, whose sign depends on the type of substrate ($D_\text{eff}<0$ and $D_\text{eff}>0$ for PM and SC, respectively), $\mathbf{n}$ is normal to the interface (Fig.~\ref{fig:1}), $\mathbf{M(q)}$ is Fourier transform of the FM film magnetization $\mathbf{M}(\bm{\rho}=x,y)$, and $\mathbf{q}=(q_x,q_y,0)$ is two-dimensional wave vector. In a frame of our model, this effective DMI depends on the system temperature $T$ through the susceptibility in PM, $\chi(T)=C/(T-T_c)$, or the London penetration depth in SC, $\lambda(T)={\lambda_0/[(1-T/T_c)]^{1/2}}$, which are contained in $D_\text{eff}$. Here $C$ and $\lambda_0$ are the PM Curie constant and the SC London penetration depth at zero temperature, respectively. The contribution (\ref{eq:1}) results from magnetostatic interaction between the film and substrate. We find that only in the case of a PM substrate, the effective DMI leads to stabilization of the Néel-type chiral magnetic textures, such as a magnetic spiral (MSp) and MSk. The strong temperature sensitivity of $D_\text{eff}$ near $T_c$ allows for tuning the effective DMI and thus controlling the MSk radius, which can be useful for the applications.

The article is organized as follows. In Section~\ref{II}, we describe the model used and derive an analytical expression for the magnetic energy of the systems under consideration. Section~\ref{III} presents conditions for stability of one-dimensional magnetic textures of the Néel-type, namely, a (1) domain wall and (2) MSp. In Section~\ref{IV}, we consider how the MSk can be stabilized. Section~\ref{V} contains a brief description of the obtained results. In Appendices~\ref{A} and \ref{B}, we calculate the magnetostatic energy in the systems under consideration and consider the case of MSp with the Bloch component.
\section{\label{II}FREE ENERGY OF THE FM/PM AND FM/SC SYSTEMS}
As for the exact geometry of our system, we place the FM film is at $0<z<h$, while a PM or SC substrate is at $z<0$ (Fig.~\ref{fig:1}). It is assumed that $h\ll L_0$, where $L_0=(2A_\text{ex}/M_0^2)^{1/2}$ is the exchange length with $A_\text{ex}$ and $M_0$ being the exchange stiffness and saturation magnetization in the FM film. As a result, the dependence of $\mathbf{M}$ on $z$ can be neglected. The nonuniform magnetization $\mathbf{M}$ induces magnetization $\mathbf{m}(\bm{\rho},z)$ or a supercurrent $\mathbf{j}_s(\bm{\rho},z)$ in the PM and SC substrates, respectively, which produce stray fields in the FM region ($z>0$). In this Section, we consider the free energy of an FM film on PM (Subsection A) and SC (Subsection B) substrates. Obtained results are generalized in Subsection C --- with distinguishing the effective DMI term.
\subsection{\label{IIA}FM/PM}
The total free energy of the FM/PM system can be written as the sum of three contributions --- $F_0$, $F_-$, and $F_+$ --- to the total free energy of the system, which are free energies within the regions $0<z<h$, $z<0$, and $z>h$, respectively, i.e.,

\begin{subequations}

\begin{equation}
F=F_0+F_-+F_+
\label{eq:2a},
\end{equation}

\begin{eqnarray}
F_0&=&\int_{-\infty}^{+\infty}\int_{0}^{h}\left[\frac{L_0^2}{2}\sum_{\alpha=1}^2\left(\frac{\partial\mathbf{M}}{\partial x_\alpha}\right)^2-\frac{1}{2}K_aM_z^2 \right.
\nonumber\\ 
& &\left.-\left(\mathbf{M}\cdot \mathbf{H}_\text{ext}\right)-\left(\mathbf{M}\cdot \mathbf{H}_0\right)-\frac{\mathbf{H}_0^2}{8\pi}\right]d\bm{\rho}\,dz
 \label{eq:2b},
\end{eqnarray}

\begin{equation}
F_-=\int_{-\infty}^{+\infty}\int_{-\infty}^{0}\left[\frac{1}{2\chi}\mathbf{m}^2-\left(\mathbf{m}\cdot\mathbf{H}_-\right)-
\frac{\mathbf{H}_-^2}{8\pi}\right]d\bm{\rho}\,dz    
\label{eq:2c},
\end{equation}

\begin{equation}
F_+=-\int_{-\infty}^{+\infty}\int_{h}^{+\infty}\frac{\mathbf{H}_+^2}{8\pi}\,d\bm{\rho}\,dz    
\label{eq:2d}.
\end{equation}

\end{subequations}
\noindent
In Eq.~(\ref{eq:2b}), the first term is nonuniform exchange, the second term is the magnetic anisotropy with $K_a>4\pi$ being the magnetic anisotropy constant, the third term is the Zeeman energy in external magnetic field $\mathbf{H}_\text{ext}$, and the fourth term relates to the magnetostatic energy. The last terms in Eqs.~(\ref{eq:2b}), (\ref{eq:2c}), and Eq.~(\ref{eq:2d}) represent the energies of the magnetic (magnetostatic) fields $\mathbf{H}_0 (0<z<h)$, $\mathbf{H}_-(z<0)$, and $\mathbf{H}_+(z>h)$ jointly produced by magnetizations $\mathbf{M}$ and $\mathbf{m}$. We assume that $\mathbf{m}=\chi\mathbf{H}_-$, which is valid under the condition that the spatial scale of $\mathbf{M}$ inhomogeneity is much larger than the exchange length in PM. Otherwise, $\mathbf{m}$ is determined by convolution of $\mathbf{H}_-$ with the integral kernel \cite{MS3}. With taking into account the electromagnetic boundary conditions \cite{Landau} and that $\mathbf{m}=\chi\mathbf{H}_-$, the total free energy reads as
\noindent
\begin{eqnarray}   
F&=&\int_{-\infty}^{+\infty}\int_{0}^{h}\left[\frac{L_0^2}{2}\sum_{\alpha=1}^2\left(\frac{\partial\mathbf{M}}{\partial x_\alpha}\right)^2-\frac{1}{2}K_a M_z^2 \right.
\nonumber\\ 
& &\left.-\left(\mathbf{M}\cdot \mathbf{H}_\text{ext}\right)-\frac{1}{2}\left(\mathbf{M}\cdot \mathbf{H}_0\right)\right]d\bm{\rho}\,dz
\label{eq:3}.
\end{eqnarray}
It can be seen that $F_-$ and the energies of the magnetostatic fields cancel each other. 

\subsection{\label{IIB}FM/SC}
As for the FM/SC system, we consider its free energy as a function of magnetic induction. Excluding such a modification, equations for the free energy of the FM/SC system are similar to those for the FM/PM one [Eqs.~(\ref{eq:2b})--(\ref{eq:2d})]:

\begin{subequations}

\begin{eqnarray}
F_0&=&\int_{-\infty}^{+\infty}\int_{0}^{h}\left[\frac{L_0^2}{2}\sum_{\alpha=1}^2\left(\frac{\partial\mathbf{M}}{\partial x_\alpha}\right)^2-\frac{1}{2}K_aM_z^2 \right.
\nonumber\\ 
& &\left.-\left(\mathbf{M}\cdot \mathbf{H}_\text{ext}\right)-\left(\mathbf{M}\cdot \mathbf{B}_0\right)+\frac{\mathbf{B}_0^2}{8\pi}\right]d\bm{\rho}\,dz
 \label{eq:4a},
\end{eqnarray}

\begin{equation}
F_-=\frac{1}{8\pi}\int_{-\infty}^{+\infty}\int_{-\infty}^{0}\left[\mathbf{B}_-^2+\lambda^2(\text{rot}\mathbf{B}_-)^2\right]d\bm{\rho}\,dz    
\label{eq:4b},
\end{equation}

\begin{equation}
F_+=\int_{-\infty}^{+\infty}\int_{h}^{+\infty}\frac{\mathbf{B}_+^2}{8\pi}\,d\bm{\rho}\,dz    
\label{eq:4c}.
\end{equation}

\end{subequations}
\noindent
The first and second terms in Eq.~(\ref{eq:4b}) are the energies of the magnetostatic field and the superconducting current, respectively. Similarly, the magnetostatic fields $\mathbf{B}_0 (0<z<h)$, $\mathbf{B}_-(z<0)$, and $\mathbf{B}_+(z>h)$ are generated jointly by magnetization $\mathbf{M}$ and supercurrent $\mathbf{j}_s$. We assume that $\mathbf{j}_s$ obeys the London equation, i.e., $\mathbf{j}_s=-c\mathbf{A}_-/(4\pi\lambda^2)$, where $\mathbf{B}_-=\text{rot}\mathbf{A}_-$. Using this equation along with the electromagnetic boundary conditions \cite{Landau}, we can reduce the free energy (\ref{eq:2a}) for the FM/SC system

\begin{eqnarray}   
F&=&\int_{-\infty}^{+\infty}\int_{0}^{h}\left[\frac{L_0^2}{2}\sum_{\alpha=1}^2\left(\frac{\partial\mathbf{M}}{\partial x_\alpha}\right)^2-\frac{1}{2}K_a M_z^2 \right.
\nonumber\\
& & \left.-\left(\mathbf{M}\cdot \mathbf{H}_\text{ext}\right)-\frac{1}{2}\left(\mathbf{M}\cdot \mathbf{B}_0\right)\right]d\bm{\rho}\,dz
\label{eq:5}.
\end{eqnarray}
We see again that $F_-$ and the energies of the magnetostatic fields cancel each other.

\subsection{\label{IIC}Magnetostatic energy and effective DMI}
The last terms in Eqs.~(\ref{eq:3}) and (\ref{eq:5}) are the magnetostatic energy, $F_\text{MS}=F_\text{MS}^\text{intra}+F_\text{MS}^\text{inter}$, which consists of two contributions. One of them, $F_\text{MS}^\text{intra}$, is the interaction between magnetic moments inside the FM film (intralayer contribution), while another one, $F_\text{MS}^\text{inter}$, is the interaction between $\mathbf{M}$ and $\mathbf{m}$ or $\mathbf{j}_s$ (interlayer contribution). The terms above can be written as follows (see Appendix~\ref{A} for their derivation):
\begin{subequations}
\begin{eqnarray}
F_\text{MS}^\text{intra(inter)}&=&\frac{S^2}{(2\pi)^2}\int_{-\infty}^{+\infty}D_{\alpha\beta}^\text{intra(inter)}(\mathbf{q}) \nonumber\\
& &\times M_\alpha(-\mathbf{q})M_\beta(\mathbf{q})\,d\mathbf{q}
\label{eq:6a},
\end{eqnarray}
\begin{eqnarray}
D_{\alpha\beta}^\text{intra}(\mathbf{q})&=&\frac{2\pi}{q}\left\{\left[qh-\left(1-e^{-qh}\right)\right]\frac{q_\alpha q_\beta}{q^2} \right. 
\nonumber \\ 
& &+\left(1-e^{-qh}\right)\delta_{\alpha z}\delta_{\beta z} \biggl \}
\label{eq:6b},
\end{eqnarray}

\begin{eqnarray}
D_{\alpha\beta}^\text{inter}(\mathbf{q})&=&qhD_\text{eff}(q)\left(\frac{q_\alpha q_\beta}{q^2}-\frac{iq_\alpha}{q}\delta_{\beta z}+ \right. 
\nonumber \\ 
& &\left.+\frac{iq_\beta}{q}\delta_{\alpha z}+\delta_{\alpha z}\delta_{\beta z}\right)
\label{eq:6c},
\end{eqnarray}

\end{subequations}
\noindent
where $S$ is the area of the system, $i$ is the imaginary unit, $\delta_{\alpha\beta}$ is the Kronecker symbol, and the effective DMI constant $D_\text{eff}$ has the form
\begin{equation}
D_\text{eff}(q)=-\frac{\pi\kappa(q)}{q^2h}\left(1-e^{-qh}\right)^2
\label{eq:7},
\end{equation}
where $\kappa(q)$ is the parameter which describes the substrate properties and depends on $T$,
\begin{equation}
\kappa(q)=\left\{
\begin{array}{rcl}
&\frac{2\pi\chi}{1+2\pi\chi}&\text{ for FM/PM,}\\

&-\frac{\sqrt{q^2\lambda^2+1}-q\lambda}{\sqrt{q^2\lambda^2+1}+q\lambda}&\text{ for FM/SC,}\\
\end{array}
\right.
\label{eq:8}
\end{equation}
which lies in the ranges of $[0,1]$ and $[-1,0]$ in FM/PM and FM/SC, respectively. Note that the sign of $\kappa$ (and hence the sign of $F_\text{MS}^\text{inter}$) depends on the substrate type and that $|\kappa|\rightarrow1$ if $T=T_c$ (FM/PM case, $\chi\rightarrow\infty$) and $T=0$ (FM/SC case, $\lambda=\lambda_0\rightarrow0$).

As the form of $F_\text{MS}^\text{intra}$ is similar to that of $F_\text{MS}^\text{inter}$, one can introduce the tensor $D_{\alpha\beta}=D_{\alpha\beta}^\text{intra}+D_{\alpha\beta}^\text{inter}$, which describes the total magnetostatic energy $F_\text{MS}$. This tensor can be represented as the sum of two terms, i.e., $D_{\alpha\beta}=D_{\alpha\beta}^s+D_{\alpha\beta}^a$, where $D_{\alpha\beta}^s=D_{\beta\alpha}^s$ is a symmetric tensor and $D_{\alpha\beta}^a=-D_{\beta\alpha}^a$ is an antisymmetric tensor whose components are nonzero only in the presence of a substrate. The $D_{\alpha\beta}^a$ tensor can be represented as 
$D_{\alpha\beta}^a=ihD_\text{eff}(q)\varepsilon_{\alpha\beta\gamma}{(\mathbf{n}\times\mathbf{q})}_\gamma$, where $\varepsilon_{\alpha\beta\gamma}$ is the Levi-Civita tensor. Substituting $D_{\alpha\beta}^a$ into Eq.~(\ref{eq:6a}), we obtain the expression for the effective DMI energy (\ref{eq:1}), where the coefficient of proportionality is $ihS^2/(2\pi)^2$. The interfacial DMI energy arising in the FM/HM system can also be reduced to the form given by Eq.~(\ref{eq:1}) after replacing $D_\text{eff}$ with a DMI constant independent of $q$.

If the spatial scale of $\mathbf{M}(\bm{\rho})$ is much larger than the film thickness $h$, one can assume that $qh\ll1$. So, expanding $D_{\alpha\beta}$ into a series up to terms linear in $qh$, one obtains that
\begin{eqnarray}
D_{\alpha\beta}&\approx&\pi h\left\{qh\left[1-\kappa(q)\right]\frac{q_\alpha q_\beta}{q^2}+2\left[1-\frac{1}{2}\kappa(q)qh\right]\delta_{\alpha z}\delta_{\beta z} \right. 
\nonumber \\ 
& &\left.+\kappa(q)qh\left(\frac{iq_\alpha}{q}\delta_{\beta z}-\frac{iq_\beta}{q}\delta_{\alpha z}\right)\right\}
\label{eq:9}.
\end{eqnarray}
If $1-\kappa\ll1$ (FM/PM case, $T\sim T_c$) one can neglect the first term in Eq.~(\ref{eq:9}). One can also neglect $\kappa qh/2$ due to $qh\ll1$ in the second term of Eq.~(\ref{eq:9}) \footnote{In fact, it is incorrect to neglect the term linear in $qh$. However, next we will make sure that neglecting this term will not lead to significant mistakes.}. As a result, the total free energy (\ref{eq:3}) can be rewritten as
\begin{eqnarray}
F&=&h\int_{-\infty}^{+\infty}\left[\frac{L_0^2}{2}\sum_{\alpha=1}^{2}{\left(\frac{\partial\mathbf{M}}{\partial x_\alpha}\right)^2-\frac{1}{2}}K_a^\text{eff}M_z^2 \right. \nonumber \\
& &\left. -\left(\mathbf{M}\cdot\mathbf{H}_\text{ext}\right)+f_\text{DMI}^\text{eff}\right]\,d\bm{\rho}
\label{eq:10},
\end{eqnarray}
where $K_a^\text{eff}=K_a-4\pi$, and
\begin{eqnarray}
f_\text{DMI}^\text{eff}&=&-\pi\kappa h\left(M_z\frac{\partial M_x}{\partial x}-M_x\frac{\partial M_z}{\partial x}\right.\nonumber \\
& &\left. +M_z\frac{\partial M_y}{\partial y}-M_y\frac{\partial M_z}{\partial y}\right)
\label{eq:11}.
\end{eqnarray}
It can be seen that, in this approximation, taking into account the magnetostatic energy leads to renormalization of the anisotropy constant and the appearance of the term $f_\text{DMI}^\text{eff}$, which has the conventional form for the DMI contribution \cite{Bog}. This energy can be evaluated as $\pi\kappa hM_0^2\sim1\text{ erg/cm}^2$ at $\pi h\sim10\text{ nm}$, $M_0\sim10^3\text{ erg/(Gs}\cdot\text{cm}^3)$, which is comparable to the DMI energy in the FM/HM system \cite{DMISW2}.
\section{\label{III}ONE-DIMENSIONAL CASE}
When FM magnetization $\mathbf{M}$ depends on a one coordinate ($x$) only, it can be represented as	
\begin{eqnarray}
\mathbf{M}(x)&=&M_0\left[\text{sin}\Theta(x)\,\text{cos}\psi\,{\hat{\mathbf{e}}}_x+\text{sin}\Theta(x)\,\text{sin}\psi\,{\hat{\mathbf{e}}}_y \nonumber \right. \\
& &\left. +\text{cos}\Theta(x)\,{\hat{\mathbf{e}}}_z\right]
\label{eq:12},
\end{eqnarray}
where ${\hat{\mathbf{e}}}_i$ is the $i$-th unit vector of the Cartesian coordinate system, $\psi$ is the constant, and $\Theta(x)$ is a function of $x$. At $\psi=\pm\pi/2$ and $\psi=0$ or $\pi$, configuration of $\mathbf{M}$ is of the Bloch- and Néel-type, respectively. We study the stabilization problem of both localized (domain wall) and delocalized (MSp) magnetic textures. Although calculations of the MSp energy in an FM film on PM and SC substrates were already reported \cite{MS1,MS5}, stabilizing such a texture has not been discussed so far. Moreover, the case of the PM substrate is considered only in the linear approximation in $\chi$.
\subsection{\label{IIIA}Domain wall}
If the boundary conditions are $\Theta(-\infty)=0$ and $\Theta(\infty)=\pi$, then Eq.~(\ref{eq:12}) describes the Néel- ($\psi=0$, $\pi$) or Bloch-type ($\psi=\pm\pi/2$) domain wall. We restrict ourselves to considering the case of a PM substrate only. After substituting Eq.~(\ref{eq:12}) into Eqs.~(\ref{eq:10}), 
 (\ref{eq:11}) and varying it with respect to $\Theta$, one obtains at $H_\text{ext}=0$ that \cite{DMI2}
\begin{equation}
\frac{d^2\Theta}{dx^2}-\frac{K_a^\text{eff}}{L_0^2}\,\text{sin}\Theta\,\text{cos}\Theta=0
\label{eq:13},
\end{equation}
whose solution for the these boundary conditions is $\Theta(x)=\pi-\text{arccos}{\left(\text{tanh}\sqrt{K_a^\text{eff}}x/L_0\right)}$. For the free energy of a film with domain wall, we have that
\begin{equation}
\Delta F=F-F_0\propto2\eta\sqrt{K_a^\text{eff}}-\pi^2\kappa\,\text{cos}\psi
\label{eq:14},
\end{equation}
where $\eta=L_0/h$, and $F_0$ is the energy of a film uniformly magnetized along the normal, i.e., $F_0=-K_a^\text{eff}M_0^2hS/2$. The obtained magnetic state exists if $\kappa$ is small enough not to perturb too much the system ($\Delta F>0$) \cite{DMI2,Bog2,Bog3}. This condition can be rewritten as $\kappa<\kappa_c^\text{DW}$ and $\psi=0$, where $\kappa_c^\text{DW}=2\eta\sqrt{K_a^\text{eff}}/\pi^2$. For a certain set of parameters, there is the interval of $\kappa$ in which ${\Delta F>0}$. For example, $\kappa_c^\text{DW}\approx0.7$ ($\chi\approx0.4$) at $\eta\sim3$ and $K_a/(4\pi)\sim1.1$.
\subsection{\label{IIIB}MSp}
Under the condition that $\Theta(x)=kx$, Eq.~(\ref{eq:12}) describes the MSp (Fig.~\ref{fig:1}). Such a trial function $\Theta(x)$ allows for calculation of the exact magnetostatic energy (\ref{eq:6a}). Having calculated $\mathbf{M}(\mathbf{q})$, one obtains the free energy, $\Delta F(k,\sigma)=F(k,\sigma)-F_0$, at $H_\text{ext}=0$ and $\psi=0$, $\pi$, i.e.,
\begin{eqnarray}
\Delta F(k,\sigma)&=&\frac{1}{2}M_0^2hS\left[L_0^2k^2+\frac{1}{2}K_a^\text{eff} \right. \nonumber \\
& &\left. -\frac{\pi\kappa(k)}{kh}(1-e^{-kh})^2(1+\sigma)^2\right]
\label{eq:15},
\end{eqnarray}
if $k\neq0$ and $\Delta F(k,\sigma)=0$ if $k=0$. Here $\sigma=1$ ($-1$) corresponds to $\psi=0$ ($\pi$); see Appendix B for arbitrary $\psi$. Equation~(\ref{eq:15}) is consistent with results of the calculations performed in Refs.~\cite{MS1,MS5} and shows that the PM (SC) substrate decreases (increases) the energy of the system. The last statement is valid in the case of an arbitrary distribution of $\mathbf{M}$ (see Appendix~\ref{A}). From Eq.~(\ref{eq:15}) we can also see that only the interlayer magnetostatic energy, which is proportional to $\kappa$, depends on the direction of MSp rotation. In the case of FM/PM(SC), the rotation corresponding to $\sigma=1$ ($-1$) is energetically favorable, which is a result of removing the chiral degeneracy of the system.  All these features can be understood using the method of images \cite{Landau} (Fig.~\ref{fig:1}). In Fig.~\ref{fig:1} the solid and dashed arrows show, respectively, the magnetic moments and their images for the cases of PM ($\kappa=1$) and SC ($\kappa=-1$) substrates. The boundary condition is vanishing of the tangential [for FM/PM, (Fig.~\ref{fig:1}a,b)] and normal [FM/SC, (Fig.~\ref{fig:1}c,d)] components of the magnetostatic field, which determine the mutual orientation of the magnetic moments and their images. The difference between the free energy in the cases of $\sigma=1$ (Fig.~\ref{fig:1}a,c) and $\sigma=-1$ (Fig.~\ref{fig:1}b,d) results from different configurations of the arrows.

Assuming that the period of the MSp is much larger than the film thickness $h$, i.e., $kh\ll1$, we can obtain expressions for the equilibrium $k^\ast$ and corresponding minimum of the free energy at $\sigma=1$,
\begin{equation}
k^\ast h\approx\frac{2\pi\kappa}{\eta^2},\ \Delta F(k^\ast,1)=\frac{1}{4}M_0^2hS\left(K_a^\text{eff}-\frac{8\pi^2\kappa^2}{\eta^2}\right)
\label{eq:16},
\end{equation}
where it was taken into account that $\eta\gg1$. Equation~(\ref{eq:16}) are valid under the condition that $\kappa>1/2$, which can only be satisfied for the FM/PM system and provides the stability of the state with respect to deviation of $\psi$ from zero (see Appendix B). From the condition that $\Delta F(k^\ast,1)=0$, one can determine the critical $\kappa$, i.e., $\kappa_c^\text{MSp}\approx\eta\sqrt{K_a^\text{eff}}/(2\sqrt2\pi)$. For $\kappa>\kappa_c^\text{MSp}$ the formation of MSp is energetically favorable. Since $\kappa\le1$, it is necessary to require the fulfillment of the relation $\kappa_c^\text{MSp}<1$. As a result, $\kappa_c^\text{MSp}\approx0.6$  ($\chi\approx0.2$) at $\eta\sim5$ and $K_a/(4\pi)\sim1.1$.

As for the FM/SC, the case of $\sigma=-1$ corresponds to $k^\ast=0$ [see Eq.~(\ref{eq:15})]. Thus, only the PM substrate stabilizes the Néel-type MSp with $\sigma=1$.
\section{\label{IV}MS\lowercase{k} STABILIZATION}
Now we consider the magnetization distribution with cylindrical symmetry (MSk) in the FM/PM case. As for the SC substrate, it increases the system energy, and so, MSk cannot be energetically favorable (see Appendix~A). The existence of a metastable MSk in the FM/SC is beyond the scope of this paper and requires additional research. 

We represent $\mathbf{M}$ as
\begin{equation}
\mathbf{M}(\rho)=M_0\left[\text{sin}\Theta(\rho)\,\hat{\mathbf{e}}_\rho+\text{cos}\Theta(\rho)\,\hat{\mathbf{e}}_z\right]
\label{eq:17},
\end{equation}
where $\hat{\mathbf{e}}_\rho=\bm{\rho}/\rho$. Equation~(\ref{eq:17}) corresponds to the configuration shown in Fig.~\ref{fig:1}a. Let the external magnetic field $\mathbf{H}_\text{ext}$ be directed against the $z$-axis, so that $\Theta(0)=0$ and $\Theta(\infty)=\pi$. As a trial function, we take the simplest linear ansatz~\cite{Bog},

\begin{equation}
\Theta(\rho)=\frac{\pi\rho}{R}\theta(R-\rho)+\pi\theta(\rho-R)
\label{eq:18},
\end{equation}
where $R$ is the MSk radius and $\theta(t)$ is the Heaviside step function. Taking into account Eqs.~(\ref{eq:17}) and (\ref{eq:18}), the approximate free energy~(\ref{eq:10}), $\Delta F(\xi)=F(\xi)-F_0$, takes the form
\begin{eqnarray}
\Delta F(\xi)&=&\pi M_0^2h^3\left\{6.15\,\eta^2+\frac{1}{16}\left[K_a^\text{eff} \right. \right. \nonumber \\
& &\left. \left.+4\left(1-\frac{4}{\pi^2}\right)\frac{H_\text{ext}}{M_0}\right]\xi^2-\frac{\pi^2\kappa}{2}\xi\right\}
\label{eq:19},
\end{eqnarray}
where $\xi=2R/h$, $F_0=-M_0^2hS(K_a^\text{eff}/2+H_\text{ext}/M_0)$. The minimum of $\Delta F(\xi)$ is provided by the competition between the anisotropy energy with $K_a^\text{eff}$ and the effective DMI energy. Note that for trial function~(\ref{eq:18}) the exchange energy does not depend on $\xi$. As a result of free energy minimization, we obtain the equilibrium parameter $\xi^\ast$ and the corresponding energy $\Delta F(\xi^\ast)$ in the form
\begin{subequations}
\begin{equation}
\xi^\ast=\frac{16\pi^2\kappa}{K_a^\text{eff}+4\left(1-4/\pi^2\right)H_\text{ext}/M_0}
\label{eq:20a}, 
\end{equation}
\begin{eqnarray}
\Delta F(\xi^\ast)&=&\pi M_0^2h^3\left[6.15\,\eta^2- \right. \nonumber \\
& & \left. \frac{16\pi^4\kappa^2}{K_a^\text{eff}+4\left(1-4/\pi^2\right)H_\text{ext}/M_0}\right]
\label{eq:20b}.
\end{eqnarray}
\end{subequations}
The obtained magnetic state is energetically favorable fore $\kappa>\kappa_c^\text{MSk}$, where $\kappa_c^\text{MSk}\approx0.06\,\eta\left(K_a^\text{eff}+2.4\,H_\text{ext}/M_0\right)^{1/2}$. So, $\kappa_c^\text{MSk}\approx0.3$ ($\chi\approx0.07$) at $\eta\sim5$, $K_a/(4\pi)\sim1.1$ and $H_\text{ext}=0$. 

\begin{figure}[h!!]
\includegraphics{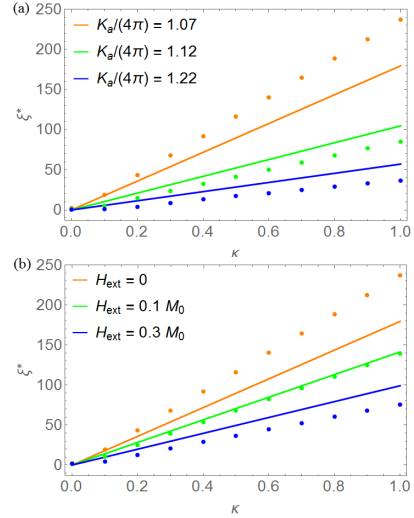}
\caption{\label{fig:2}Equilibrium parameter $\xi^\ast$ vs. $\kappa$ in FM/PM for different (a) $K_a$ and (b) $H_\text{ext}$. The solid lines show the calculations using the approximate Eq.~(\ref{eq:20a}), the dots show the numerical calculations based on the exact Eqs. (\ref{eq:A18})--(\ref{eq:A19}), (\ref{eq:21a})--(\ref{eq:21c}).}
\end{figure}

We also present numerical calculations based on the exact magnetostatic energy $F_\text{MS}$. Then $F_\text{MS}^\text{intra}$ and $F_\text{MS}^\text{inter}$ are described by Eqs.~(\ref{eq:A18}) and (\ref{eq:A19}), where	
\begin{subequations}
\begin{equation}
M_z(\mathbf{q})=\Delta M_z(q)-\frac{(2\pi)^2}{S}M_0\,\delta(\textbf{q})
\label{eq:21a},
\end{equation}
\begin{equation}
\Delta M_z(q)=\frac{2\pi M_0}{S}\int_{0}^{\infty}\rho\left[\text{cos}\Theta(\rho)+1\right]J_0(q\rho)\,d\rho
\label{eq:21b},
\end{equation}
\begin{eqnarray}
\text{div}\textbf{M}(q)&=&\frac{2\pi M_0\sigma}{S}\int_{0}^{\infty}\left[\rho\frac{d\Theta(\rho)}{d\rho}\text{cos}\Theta(\rho) \right. \nonumber \\
& & \left. +\text{sin}\Theta(\rho)\right]J_0(q\rho)\,d\rho
\label{eq:21c}.
\end{eqnarray}
\end{subequations}
Here $\text{div}\mathbf{M}(\mathbf{q})=\text{div}\mathbf{M}(q)$ is the Fourier transform of $\text{div}\mathbf{M}(\bm{\rho})$; $\delta(\mathbf{q})$ and $J_n(t)$ are the Dirac delta function and cylindrical Bessel function of order $n$. Figure~2 shows the parameter $\xi^\ast$ versus $\kappa$ for various $K_a$ (Fig.~\ref{fig:2}a) and $H_\text{ext}$ (Fig.~\ref{fig:2}b). In Fig.~\ref{fig:2}a,b the solid lines correspond to approximate Eq.~(\ref{eq:20a}), while the dots correspond to numerical calculations based on exact Eqs.~(\ref{eq:A18})--(\ref{eq:A19}), (\ref{eq:21a})--(\ref{eq:21c}). It can be seen that the Eq.~(\ref{eq:20a}) agrees with the exact calculation. The difference in the results seems to be due to the neglect of $\kappa qh/2$ in the second term of Eq.~(\ref{eq:9}). Nevertheless, the obtained approximate results are quite suitable for our evaluations.

Thus, MSk can be energetically favorable in FM/PM, which is in contrast to that in the FM/SC.
\section{\label{V}SUMMARY}
In conclusion, we have shown that the effective DMI term having a purely magnetostatic origin contributes to the free energy of the FM/PM and FM/SC hybrid systems. The value of this term can be evaluated as $\pi\kappa hM_0^2\sim1\text{ erg/cm}^2$ at $\pi h\sim10$ nm, $M_0\sim10^3 \text{erg}/(\text{Gs}\cdot\text{cm}^3)$, which is comparable to the conventional DMI energy in the FM/HM system. We have shown that the sign of the effective DMI constant is different for PM and SC substrates, i.e., $D_\text{eff}<0$ ($D_\text{eff}>0$) for FM/PM(SC). The value of $D_\text{eff}$ strongly depends on the system temperature near $T_c$, which allows for tuning the interaction between the film and substrate. Note that $D_\text{eff}$ can be measured experimentally by Brillouin spectroscopy \cite{DMISW2}. We have also shown that the effective DMI leads to stabilizing chiral magnetic textures (including MSk) only in the case of a PM substrate. This feature is explicable by decreasing the free energy of FM film in the FM/PM system owing to the effect of PM substrate, while it only increases in the FM/SC system. The difference between FM/PM and FM/SC systems is related to the difference in the formation of image dipoles in the PM and SC substrates (Fig.~\ref{fig:1}). To stabilize the MSk at $H_\text{ext}=0$, the condition that $\kappa>0.3$ ($\chi>0.07$) at $\eta\sim5$ and $K_a/(4\pi)\sim1.1$ should be fullfilled. If, for example, gadolinium ($C\approx0.4$ K, $T_c\approx294$ K \cite{Gd}) is taken as PM in the FM/PM system, the condition above will be effective in the temperature range $T-T_c<6$ K.
\begin{acknowledgments}
We acknowledge very useful discussions with M.V.~Sapozhnikov, A.S.~Mel’nikov, and I.S.~Burmistrov. The work was supported by the State Contract No.~0030-2021-0021 and Russian Foundation for Basic Research (Grant No.~20-02-00356). One of us (M.A.~K.) thanks to the Foundation for the Advancement of Theoretical Physics and Mathematics “BASIS”.
\end{acknowledgments}

\appendix

\section{\label{A}Magnetostatic energy of FM/PM and FM/SC systems}
In this Appendix, we calculate the magnetostatic energy of the FM/PM (Subsection 1)  and FM/SC (Subsection 2) systems. Obtained results are generalized in Subsection 3.
\subsection{\label{app:A1}FM/PM}
Let $\mathbf{H}$ be the magnetostatic field jointly produced by magnetizations $\mathbf{M}$ and $\mathbf{m}$,
\begin{equation}
\textbf{H}=\left\{
\begin{array}{rcl}
&\textbf{H}_-,\, z<0,\\
&\textbf{H}_0,\, 0<z<h,\\
&\textbf{H}_+,\, z>0.\\
\end{array}
\right.
\label{eq:A1}
\end{equation}
We introduce a scalar potential $\varphi$, so that $\mathbf{H}=-\nabla\varphi$. Then the Maxwell's equations, $\text{div}\mathbf{B}=0$ and $\text{rot}\mathbf{H}=0$, where $\mathbf{B}=\mathbf{H}+4\pi\mathbf{\mathfrak{M}}$, $\mathbf{\mathfrak{M}}(\bm{\rho},z)=\mathbf{m}(\bm{\rho},z)\theta(-z)+\mathbf{M}(\bm{\rho})\theta(z)\theta(h-z)$, are converted to the Poisson equation $\Delta\varphi=4\pi \text{div}\mathbf{\mathfrak{M}}$. We can write its solution in terms of the magnetic field,
\begin{equation}
\mathbf{H}=\mathbf{H}^M+\mathbf{H}^m
\label{eq:A2},
\end{equation}
where $\mathbf{H}^M$ and $\mathbf{H}^m$ are contributions from the magnetization of the film and substrate, respectively,
\begin{eqnarray}
\mathbf{H}^M(\bm{\rho},z)&=&-\hat{\mathbf{e}}_\alpha\int_{-\infty}^{+\infty}\int_{0}^{h}D_{\alpha\beta}^\text{intra}(\bm{\rho}-\bm{\rho}^\prime,z-z^\prime) \nonumber \\
& &\times M_\beta(\bm{\rho}^\prime)\,d\bm{\rho}^\prime\,dz
\label{eq:A3},
\end{eqnarray}
\begin{equation}
\mathbf{H}^m(\bm{\rho},z)=-\int_{-\infty}^{+\infty}m_z(\bm{\rho}^\prime,0)\nabla\frac{1}{|\mathbf{r}-\mathbf{r}^\prime|}|_{z^\prime=0}\,d\bm{\rho}^\prime
\label{eq:A4}.
\end{equation}
Here $\mathbf{r}=(\bm{\rho},z)$, $\mathbf{r}^\prime=(\bm{\rho}^\prime,z^\prime)$, $\hat{\mathbf{e}}_\alpha$ --- $\alpha$-th unit vector of the Cartesian coordinate system, and $D_{\alpha\beta}^\text{intra}$ is symmetric tensor of the second rank,
\begin{equation}
D_{\alpha\beta}^\text{intra}(\bm{\rho}-\bm{\rho}^\prime,z-z^\prime)=\frac{\partial^2}{\partial x_\alpha\partial x_\beta^\prime}\frac{1}{\left|\mathbf{r}-\mathbf{r}^\prime\right|}
\label{eq:A5}.
\end{equation}
From Eqs.~(\ref{eq:A2})-(\ref{eq:A4}) it is also easy to obtain an integral equation for $m_z(\bm{\rho},0)$,
\begin{eqnarray}
\frac{1}{\chi}m_z(\mathbf{\rho},0)&+&\int_{-\infty}^{+\infty}{m_z(\bm{\rho}^\prime,0)\left(\frac{\partial}{\partial z}\frac{1}{\left|\mathbf{r}-\mathbf{r}^\prime\right|}\right)|_{z=0}d\bm{\rho}^\prime} \nonumber \\
 &=&\left[\mathbf{n}\cdot\mathbf{H}_-^M(\bm{\rho},0)\right]
\label{eq:A6}.
\end{eqnarray}
Solving Eq.~(\ref{eq:A6}) allows to define the field $\mathbf{H}$. Finally, we can write expressions for $F_\text{MS}^\text{intra}$ and $F_\text{MS}^\text{inter}$,
\begin{eqnarray}
F_\text{MS}^\text{intra}&=&-\frac{1}{2}\int_{-\infty}^{+\infty}\int_{0}^{h}(\mathbf{M}\cdot\mathbf{H}_0^M)\,d\bm{\rho}\,dz \nonumber \\
&=&\frac{1}{2}\iint_{-\infty}^{+\infty}\iint_{0}^{h}{D_{\alpha\beta}^\text{intra}\left(\bm{\rho}-\bm{\rho}^\prime,z-z^\prime\right)} \nonumber \\
& & \times M_\alpha\left(\bm{\rho}\right)M_\beta\left(\bm{\rho}^\prime\right)\,d\bm{\rho}\,d\bm{\rho}^\prime\, dz\,dz^\prime 
\label{eq:A7},
\end{eqnarray}
\begin{eqnarray}
F_\text{MS}^\text{inter}&=&-\frac{1}{2}\int_{-\infty}^{+\infty}\int_{0}^{h}(\mathbf{M}\cdot\mathbf{H}_0^m)\,d\bm{\rho}\,dz \nonumber \\
&=&\frac{1}{2}\iint_{-\infty}^{+\infty}\int_{0}^{h}m_z\left(\bm{\rho}^\prime,0\right)\left[\mathbf{M}\left(\bm{\rho}\right)\cdot\nabla\right] \nonumber \\
& &\times\frac{1}{\left|\mathbf{r}-\mathbf{r}^\prime\right|}|_{z^\prime=0}\,d\bm{\rho}\,d\bm{\rho}^\prime\,dz
\label{eq:A8}.
\end{eqnarray}
\subsection{\label{app:A2}FM/SC}
Similarly, we introduce the magnetostatic field $\mathbf{B}$ jointly produced by $\mathbf{M}$ and $\mathbf{j}_s$,	
\begin{equation}
\textbf{B}=\left\{
\begin{array}{rcl}
&\textbf{B}_-,\, z<0,\\
&\textbf{B}_0,\, 0<z<h,\\
&\textbf{B}_+,\, z>0,\\
\end{array}
\right.
\label{eq:A9}
\end{equation}
so $\mathbf{B}=\text{rot}\mathbf{A}$, where $\mathbf{A}$ is a vector potential. Using the Maxwell's equations and the London one, i.e., $\text{rot}\mathbf{B}=4\pi\mathbf{j}/c$, $\text{div}\mathbf{B}=0$, $\mathbf{j}(\bm{\rho},z)=c\,\text{rot}[\mathbf{M}(\bm{\rho})\theta(z)\theta(h-z)]+\theta(-z)\mathbf{j}_s(\bm{\rho},z)$, $\mathbf{j}_s=-c\mathbf{A}_-/(4\pi\lambda^2)$, we obtain the equation for the vector potential
\begin{eqnarray}
\Delta \mathbf{A}&=&-4\pi\left\{\theta(z)\theta(h-z)\text{rot}\mathbf{M} \right. \nonumber \\
& & \left. +\left[\delta(z)-\delta(h-z)\right](\mathbf{n}\times\mathbf{M}) \right. \nonumber \\
& &\left. +\frac{1}{c}\theta(-z)\mathbf{j}_s\right\}
\label{eq:A10}.
\end{eqnarray}
Note that the London equation requires the introduction of the London gauge, i.e., $\text{div}\mathbf{A}=0$ and $(\mathbf{n}\cdot\mathbf{A}_-)|_{z=0}=0$. The solution of Eq.~(\ref{eq:A10}) has the form
\begin{equation}
\mathbf{A}=\mathbf{A}^\ast+\mathbf{A}^M+\mathbf{A}^s
\label{eq:A11},
\end{equation}
where $\mathbf{A}^M$ and $\mathbf{A}^s$ are the contributions from the film magnetization and the substrate supercurrent, respectively,
\begin{equation}
\mathbf{A}^M(\bm{\rho},z)=\int_{-\infty}^{+\infty}\int_{0}^{h}\left[\mathbf{M}(\bm{\rho}^\prime)\times\nabla^\prime\frac{1}{\left|\mathbf{r}-\mathbf{r}^\prime\right|}\right]d\bm{\rho}^\prime dz^\prime
\label{eq:A12},
\end{equation}
\begin{equation}
\mathbf{A}^s(\bm{\rho},z)=\frac{1}{c}\int_{-\infty}^{+\infty}\int_{-\infty}^{0}{\frac{1}{\left|\mathbf{r}-\mathbf{r}^\prime\right|}\mathbf{j}_s(\bm{\rho}^\prime,z^\prime)d\bm{\rho}^\prime dz^\prime}
\label{eq:A13}.
\end{equation}
Here the operator $\nabla^\prime$ acts on the $\mathbf{r}^\prime$ coordinates, and the term $\mathbf{A}^\ast$ is the solution of the homogeneous equation $\Delta \mathbf{A}^\ast=0$. The introduction of $\mathbf{A}^\ast$ ensures the validity of the London gauge. From Eqs.~(\ref{eq:A11})-(\ref{eq:A13}) one can obtain an integral equation for $\mathbf{j}_s$,
\begin{eqnarray}
\mathbf{j}_s(\bm{\rho},z)&+&\frac{1}{4\pi\lambda^2}\int_{-\infty}^{+\infty}\int_{-\infty}^{0}{\frac{1}{\left|\mathbf{r}-\mathbf{r}^\prime\right|}\mathbf{j}_s(\bm{\rho}^\prime,z^\prime)d\bm{\rho}^\prime d z^\prime} \nonumber \\
&=&-\frac{c}{4\pi\lambda^2}\left[\mathbf{A}_-^M(\bm{\rho},z)+\mathbf{A}_-^\ast(\bm{\rho},z)\right]
\label{eq:A14},
\end{eqnarray}
whose solution will allow us to determine $\mathbf{B}$. Calculating $\mathbf{B}^M=\text{rot}\mathbf{A}^M$, we obtain the equation $\mathbf{B}^M=\mathbf{H}^M+4\pi\theta(z)\theta(h-z)\mathbf{M}$, where $\mathbf{H}^M$ is defined by Eq.~(\ref{eq:A3}). Then for the energy $F_\text{MS}^\text{intra}$, after discarding the insignificant constant term proportional to the mean square of $\mathbf{M}$ and related to the different definitions of energies (\ref{eq:2b}) and (\ref{eq:4a}), we obtain Eq.~(\ref{eq:A7}). For $F_\text{MS}^\text{inter}$ we get the following expression
\begin{eqnarray}
F_\text{MS}^\text{inter}&=&-\frac{1}{2}\int_{-\infty}^{+\infty}\int_{0}^{h}\left(\mathbf{M}\cdot \text{rot}\mathbf{A}^s\right)d\bm{\rho}\,dz \nonumber \\
&=& -\frac{1}{2c}\iint_{-\infty}^{+\infty}\int_{0}^{h}\int_{-\infty}^{0}\left[\mathbf{M}(\bm{\rho}) \right. \nonumber \\
& & \left. \cdot\ \text{rot}\left(\frac{\mathbf{j}_s(\bm{\rho}^\prime,z^\prime)}{\left|\mathbf{r}-\mathbf{r}^\prime\right|}\right)\right]d\bm{\rho}\,d\bm{\rho}^\prime dz\,dz^\prime
\label{eq:A15}.
\end{eqnarray}
\subsection{\label{app:A3}Magnetostatic energy of FM/PM and FM/SC systems}
Finally, after applying the Fourier transform for the magnetization $\mathbf{M}(\bm{\rho})$,
\begin{equation}
\mathbf{M}(\bm{\rho})=\frac{S}{{(2\pi)}^2}\int_{-\infty}^{+\infty}{\mathbf{M}(\mathbf{q})e^{i(\mathbf{q}\cdot\bm{\rho})}d\mathbf{q}}
\label{eq:A16},
\end{equation}
\begin{equation}
\mathbf{M}(\mathbf{q})=\frac{1}{S}\int_{-\infty}^{+\infty}{\mathbf{M}(\bm{\rho})e^{-i(\mathbf{q}\cdot\bm{\rho})}d\bm{\rho}}
\label{eq:A17},
\end{equation}
as well as integration over $z$ and $z^\prime$ from Eqs.~(\ref{eq:A6})–(\ref{eq:A8}), and Eqs.~(\ref{eq:A14})–(\ref{eq:A15}) we obtain Eqs.~(\ref{eq:6a})--(\ref{eq:6c}).

Now we show that the PM substrate decreases the system energy, while the SC substrate increases it. Noting that $\mathbf{M}(-\mathbf{q})=\mathbf{M}^\ast(\mathbf{q})$ and $\text{div}\mathbf{M}(\mathbf{q})=i\left[\mathbf{q}\cdot\mathbf{M}(\mathbf{q})\right]$, where $\text{div}\mathbf{M}(\mathbf{q})$ is the Fourier transform of $\text{div}\mathbf{M}(\bm{\rho})$, we bring the energies~(\ref{eq:6a}) to the forms
\begin{eqnarray}
F_\text{MS}^\text{intra}&=&\frac{S^2}{2\pi}\int_{-\infty}^{+\infty}\frac{1}{q}\left\{\frac{1}{q^2}\left[qh-\left(1-e^{-qh}\right)\right]\left|\text{div}\mathbf{M}(\mathbf{q})\right|^2 \right. \nonumber \\
& & \left. +\left(1-e^{-qh}\right)\left|M_z(\mathbf{q})\right|^2\right\}d\mathbf{q}
\label{eq:A18},
\end{eqnarray}
\begin{equation}
F_\text{MS}^\text{inter}=\frac{hS^2}{\left(2\pi\right)^2}\int_{-\infty}^{+\infty}{\frac{1}{q}D_\text{eff}(q)}\left|\text{div}\mathbf{M}(\mathbf{q})+qM_z(\mathbf{q})\right|^2d\mathbf{q}
\label{eq:A19}.
\end{equation}
Equation~(\ref{eq:A19}) shows that the sign of $F_\text{MS}^\text{inter}$ is determined by the sign of $D_\text{eff}$. Since $D_\text{eff}<0$ ($D_\text{eff}>0$) for FM/PM(SC), the PM(SC) substrate decreases (increases) the energy of the system.
\section{\label{B}MSp with an arbitrary $\psi$}
Here we consider the case of the MSp with an arbitrary $\psi$ to determine the conditions under which a state with $\psi=0$ is stable with respect to a change in $\psi$. After calculating $\mathbf{M}(\mathbf{q})$ for $\Theta(x)=kx$ and substituting it into Eq.~(\ref{eq:6a}), we get the expression for the free energy $\Delta F(k,\psi)=F(k,\psi)-F_0$ at $H_\text{ext}=0$, i.e.,	
\begin{eqnarray}
\Delta F(k,\psi)&=&M_0^2hS\left(\frac{1}{2}L_0^2k^2+\frac{1}{4}K_a^\text{eff} \right. \nonumber \\
& & \left.
-\frac{\pi}{kh}\left\{\left[kh-\left(1-e^{-kh}\right)\right]\text{sin}^2\psi \right. \right. \nonumber \\
& & \left. \left. +\frac{\kappa(k)}{2}\left(1-e^{-kh}\right)^2\left(1+\text{cos}\psi\right)^2\right\}\right)
\label{eq:B1}
\end{eqnarray}
if $k\neq0$, and $\Delta F=0$ if $k=0$. Note that the "Bloch contribution" to the magnetization (\ref{eq:12}) does not depend on the sign of $\psi$. This circumstance is connected with the disappearance of volume magnetic charges at $\psi=\pm\pi/2$, i.e., $\text{div}\mathbf{M}=0$. Assuming that $kh\ll1$, we obtain expressions for the equilibrium $k^\ast$ and corresponding minimum of the free energy,
\begin{equation}
k^\ast=\frac{\pi}{2h\eta^2}\left[\text{sin}^2\psi+\kappa\left(1+\text{cos}\psi\right)^2\right]
\label{eq:B2},
\end{equation}
\begin{eqnarray}
\Delta F(k^\ast,\psi)&=&\frac{1}{4}M_0^2hS\left\{K_a^\text{eff} -\frac{\pi^2}{2\eta^2}\left[\text{sin}^2\psi \right. \right. \nonumber \\
& &\left. \left. +\kappa\left(1+\text{cos}\psi\right)^2\right]^2\right\}
\label{eq:B3},
\end{eqnarray}
where it was taken into account that $\eta\gg1$. Equations (\ref{eq:B2}) and (\ref{eq:B3}) are valid under the condition $\text{sin}^2\psi+\kappa\left(1+\text{cos}\psi\right)^2\geq0$, which is always satisfied for $\kappa\geq0$ (FM/PM case). Expanding $\Delta F(k^\ast,\psi)$ into a series in $\psi$ around $\psi=0$ and requiring that the coefficient at $\psi^2$ be positive, we obtain the conditions for the existence of a minimum at $\psi=0$, i.e., $\kappa>1/2$. Since for FM/PM $0\le\kappa\le1$,  this condition can be satisfied.

Note that in the absence of a substrate $(\kappa=0)$ the minimum of $\Delta F$ corresponds to the Bloch-type MSp. Apparently, such a solution is connected with the choice of a trial function, which leads to the independence of the anisotropy energy from $k$.

\nocite{*}

\bibliography{apssamp}

\end{document}